\documentstyle[12pt,epsf]{article}
\begin{document}
\begin{titlepage}
\begin{flushright}
\end{flushright}
\vspace{0.5cm}
\begin{center}
{\Large \bf 
$W(E_{10})$ Symmetry, $M$-Theory and \\
Painlev\'e Equations
\par}
\vskip1cm
\normalsize
{\large Shun'ya Mizoguchi}\\
\vskip 1.5em
{\it Institute of Particle and Nuclear Studies\\
High Energy Accelerator Research Organization (KEK)\\
Oho 1-1, Tsukuba, Ibaraki, 305-0801, Japan}
\vskip 1.5em
and\\
\vskip 1em
{\large Yasuhiko Yamada}
\vskip 1.5em
{\it Department of Mathematics, Kobe University \\
Rokko, Kobe 657-8501, Japan}
\vskip 3em
{\it Dedicated to the memory of Sung-Kil Yang}
\end{center}
\vskip 3em
\begin{abstract}
The Weyl group symmetry $W(E_{k})$ is studied from the points of view
of the $E$-strings, Painlev\'e equations and $U$-duality.
We give a simple reformulation of the elliptic Painlev\'e
equation in such a way that the hidden symmetry $W(E_{10})$
is manifestly realized. This reformulation is based on the
birational geometry of the del Pezzo surface and closely
related to Seiberg-Witten curves describing the $E$-strings.
The relation of the $W(E_{k})$ symmetry
to the duality of $M$-theory on a torus is discussed
on the level of string equations of motion.
\end{abstract}
\end{titlepage}
\def\be{\begin{equation}}
\def\ee{\end{equation}}
\def\ba{\begin{eqnarray}}
\def\ea{\end{eqnarray}}
\def\dfrac#1#2{{\displaystyle\frac{#1}{#2}}}

\section{Introduction}

In a recent paper \cite{Sakai},
the second order (difference) Painlev\'e equations have been
classified by using the geometry of algebraic surfaces.
The classification falls into three types:
rational, trigonometric and elliptic.
Each case is associated with a special divisor
corresponding to one of the Kodaira singular fibers
of elliptic fibration (Table. \ref{kunihiko}) \cite{Kod}.
\def\rI{{\rm I}}
\def\rII{{\rm II}}
\def\rIII{{\rm III}}
\def\rIV{{\rm IV}}
\renewcommand{\arraystretch}{1.2}
\begin{table}[b]
\begin{center}
\begin{tabular}{||c|c||} \hline
fiber type & singularity \\
\hline
$\rI_n \ (n \geq 1)$ &
$A_{n-1}$ \\
$\rII, \rIII, \rIV \ (n=0,1,2)$  &
$A_n$ \\
$\rI_n^* \ (n \geq 0)$ &
$D_{n+4}$ \\
$\rII^*,\rIII^*,\rIV^*  \ (n=8,7,6)$ &
$E_n$ \\
\hline
\end{tabular}
\end{center}
\caption{Kodaira's classification and ADE singularities.}
\label{kunihiko}
\end{table}
\def\rr{\hskip-3mm \rightarrow \hskip-3mm}
\def\ne{\hskip-3mm \nearrow \hskip-3mm}
\def\se{\hskip-3mm \searrow \hskip-3mm}
\def\dw{\hskip-3mm \downarrow \hskip-3mm}
\def\hE{\widehat{E}}
\be
\begin{array}{cccccccccccccccccccccccc}
{\rm ell.}&\rI_0\\[2mm]
{\rm tri.}&\rI_1&\rr&\rI_2&\rr&\rI_3&\rr&\rI_4&\rr&\rI_5&\rr&\rI_6&\rr
&\rI_7&\rr&\rI_8&&\rI_9\\
&&&&&&&&&&&&&&\se&&\ne\\
&&&&&&&&&&&&&&&\rI_8\\[2mm]
{\rm rat.}&\rII&\rr&\rIII&\rr&\rIV&&\rr&&
{\rI_0}^{\ast}&\rr&{\rI_1}^{\ast}&\rr
&{\rI_2}^{\ast}&\rr&{\rI_3}^{\ast}&\rr&{\rI_4}^{\ast} \\
&&&&&&&&&&&&\se&&\se&&&\dw \\
&&&&&&&&&&&&&\rIV^{\ast}&\rr&\rIII^{\ast}&\rr&\rII^{\ast}
\end{array}
\ee
In physics, the same\footnote{The corresponding root systems are
complement with each other.} diagram appeared as the RG flow
of the $E$-strings in dimensions $d=6$ \cite{GH,SW6d}, $d=5$ \cite{Sib,YY}
and $d=4$ \cite{NTY}.
\be
\begin{array}{cccccccccccccccccccccccc}
{d=6}&E_{10}\\[2mm]
{d=5}&\hE_8&\rr&\hE_7&\rr&\hE_6&\rr&\hE_5&\rr&\hE_4&\rr&\hE_3&\rr&\hE_2
&\rr&\hE_1&&\hE_0\\
&&&&&&&&&&&&&&\se&&\ne\\
&&&&&&&&&&&&&&&\widetilde{\hE}_1\\[2mm]
{d=4}&E_8&\rr&E_7&\rr&E_6&&\rr&&D_4&\rr&D_3&\rr
&D_2&\rr&D_1&\rr&D_0 \\
&&&&&&&&&&&&\se&&\se&&&\dw \\
&&&&&&&&&&&&&A_2&\rr&A_1&\rr&A_0
\end{array}
\ee
Of course, this is not an accident, since both of them are described by the
same geometry, namely the del Pezzo surface $B_9$ which is
a blown up of ${\bf P}^2$ at $9$ points. The difference is that
in the $E$-string case the 9 points are chosen in special position
so that the surface admits an elliptic fibration.

Furthermore, the $E_n$-series in the above diagram is also well-known in
connection with the duality symmetry of $M$-theory compactified on a
torus \cite{Julia}. In this correspondence, the Weyl group part of the
$U$-duality group is identified with the Cremona isometry $W(E_k)$ for
del Pezzo $B_k$ \cite{INV}. 
(This duality was discussed from the point of view of Little String
Theory supersymmetric indices \cite{Bonelli}.)  The aim of this paper is
to examine the correspondence by closely
looking the way how the Weyl group is realized in each case.

This paper is organized as follows.
In section 2, we clarify the special role of the fiber at $u=\infty$
in the Seiberg-Witten geometry \cite{AKSST}.
We also give an example of duality map between two Seiberg-Witten curves
corresponding to different space-time dimensions.
In section 3, we give a reformulation of the elliptic Painlev\'e equation
where the hidden $W(E_{10})$ symmetry is manifestly realized and
the relation to the Seiberg-Witten geometry of $E$-strings is discussed.
In section 4, we study the Painlev\'e equations arising from a consistent
truncation/reduction of the $M$-theory and compare the
Painlev\'e B\"acklund transformations with the $U$-duality.
Finally, section 5 is devoted to the conclusions and discussions.

\section{The role of the fiber at $u=\infty$}

\def\sn{{\rm sn}}
\def\cn{{\rm cn}}
\def\dn{{\rm dn}}
The equations for the $SU(2)$ $E_8$ flavor Seiberg-Witten curves with
two mass parameters have been given by Minahan et.~al.\cite{Min}
\begin{equation}\label{2mass}
\begin{array}{ll}
{\rm rat.}&y^2=x^3-2 u(u^2+m_1^2 x)(u^2+m_2^2 x),\\[2mm]
{\rm tri.}&
y^2=x^3+u^2 x^2-2 u(u^2+\sin^2m_1 x)(u^2+\sin^2m_2 x),\\[2mm]
{\rm ell.}&y^2=x^3+(1+k^2)u^2 x^2-2 u(u^2+\sn^2m_1 x)(u^2+\sn^2m_2 x)
+k^2 u^4 x.
\end{array}
\end{equation}
The discriminants and singular fibers are
\begin{equation}
\begin{array}{lll}
{\rm rat.}&\Delta=u^8(u^2+\cdots),&
\rI_2^{\ast}+2\rI_1+(\rII)_{u=\infty},\\[2mm]
{\rm tri.}&\Delta=u^8(u^3+\cdots),&
\rI_2^{\ast}+3\rI_1+({\rI_1})_{u=\infty},\\[2mm]
{\rm ell.}&\Delta=u^8(u^4+\cdots),&
\rI_2^{\ast}+4\rI_1.
\end{array}
\end{equation}
The $\rI^{\ast}_2(=D_6)$ singularity at
$u=0$ corresponds to the two mass deformation of $\rII^{\ast}(=E_8)$.
Note that the difference among the three cases (rat/tri/ell)
appears on the fiber at $u=\infty$. That is,
the fiber is a cusp/nodal/smooth curve, respectively.

To see the meaning of the fiber at $u=\infty$, let us consider the
sections. For all three cases, the Mordell-Weil lattice are
$A_1^{\ast} \oplus A_1^{\ast}$.\cite{FYY}
In fact, we have the following generators of the sections,\cite{Min}
\begin{equation}\label{sec-gen}
\begin{array}{lll}
{\rm rat.}& x=-\dfrac{1}{v^2}u^2& y=i\dfrac{1}{v^3}u^3,\\[3mm]
{\rm tri.}& x=-\dfrac{1}{\sin^2v}u^2& y=i\dfrac{\cos
v}{\sin^3v}u^3,\\[3mm]
{\rm ell.}& x=-\dfrac{1}{\sn^2v}u^2& y=i\dfrac{\cn v \dn v}{\sn^3v}u^3,
\end{array}
\end{equation}
where $v=m_1$ or $m_2$.
Other sections can be obtained by addition and have the form
at $u=\infty$ as
\begin{equation}
x=a_2u^2+a_1u+\cdots, \quad
y=b_3u^2+b_2u^2+\cdots,
\end{equation}
where the leading term is of the form (\ref{sec-gen})
with $v=k_1 m_1+k_2 m_2$ ($k_1, k_2 \in {\bf Z}$).
Let us consider the elliptic case.
The fiber at $u=\infty$ is a smooth curve
\begin{equation}
y^2=x^3+(1+k^2)x^2+k^2x,
\end{equation}
which can be parametrized as
\begin{equation}
x=x(v)=-\dfrac{1}{\sn^2 v}, \quad
y=y(v)=i \dfrac{\cn v \dn v}{\sn^3 v}.
\end{equation}
Note that this is nothing but the leading term of the section.
Hence the parameter $v$ represent the point where
the section and the fiber at $u=\infty$ intersect.
Similarly, for the case of rational or trigonometric,
the intersection point is parametrized by the
trigonometric or rational functions of parameter $v$. 
In \cite{NTY} the coincidence between certain parameter $v$ in sections
and mass parameters was observed. The above argument explains the
mechanism of this identification.

Finally, we consider the relation between
the curves in (\ref{2mass}) and $SU(2)$ $N_f=2$ Seiberg-Witten curve
\cite{SW}. The $N_f=2$ Seiberg-Witten curve
\begin{equation}
y^2=(x^2-\dfrac{\Lambda^4}{64})(x-u)+\dfrac{\Lambda^2}{4}M_1 M_2 x
-\dfrac{\Lambda^4}{64}(M_1^2+M_2^2),
\end{equation}
and the elliptic case in (\ref{2mass})
are both the generic curves with the $D_6$ singularity.
Hence, they should be related with each other.
In fact, up to simple change of variables $x,u$,
these curves are equivalent. The relations of parameters are
\begin{equation}
\begin{array}c
\Lambda^2=4(\dfrac{1}{\sn m_1}-\dfrac{1}{\sn m_2}), \\[4mm]
\Lambda^2(M_1+M_2)=8\dfrac{\cn m_1 \dn m_1}{\sn^3 m_1}, \quad
\Lambda^2(M_1-M_2)=8\dfrac{\cn m_2 \dn m_2}{\sn^3 m_2}.
\end{array}
\end{equation}
This mapping $(\Lambda, M_1, M_2) \leftrightarrow (k, m_1, m_2)$
can be interpreted as a kind of duality which connects different theories
(in different dimensions).

\section{The elliptic Painlev\'e equation}

On a del Pezzo surface $B_k$ the Weyl group $W(E_k)$ acts as the
Cremona isometry \cite{OD}.
For the case of $k=9$, the Weyl group $W(E_9)$ is the affine
Weyl group of type $W(E^{(1)}_8)$ which contains the
translation subgroup ${\bf Z}^8$ and this is the origin of
the elliptic Painlev\'e equation \cite{Sakai}.
This construction can be considered as an example of general
strategy to construct discrete Painlev\'e equations by using
affine Weyl groups \cite{NY}.

We will reformulate the elliptic Painlev\'e equation
in the form where the hidden $W(E_{10})$ symmetry is manifestly realized.
Let $M$ be the space of $3 \times 10$ matrix
\be
M=\left\{
X=\left[\begin{array}{ccccc}
x_{1}&x_{2}&x_{3}&\cdots&x_{10}\\
y_{1}&y_{2}&y_{3}&\cdots&y_{10}\\
z_{1}&z_{2}&z_{3}&\cdots&z_{10}
\end{array}\right]
\right\}.
\ee
Each column vector $P_i=(x_{i}:y_{i}:z_{i})$ may be thought of as
a projective coordinate of a point $P_i \in {\bf P}^2$.
In view of this, we make an identification
\be
{\cal M}={\rm PGL}(3) \backslash M / ({\bf C}^{\times})^{10}.
\ee
A representative of this coset can be taken as
\be
X=\left[\begin{array}{cccccccc}
1&&&1&1&\cdots&1\\
&1&&1&u_5&\cdots&u_{10}\\
&&1&1&v_5&\cdots&v_{10}
\end{array}\right],
\ee
where
\be
u_i=\dfrac{\mu_{234}\mu_{13i}}{\mu_{134}\mu_{23i}}, \quad
v_i=\dfrac{\mu_{234}\mu_{12i}}{\mu_{124}\mu_{23i}}  \quad
(i=5,\ldots, 10),
\ee
and $\mu_{ijk}$ is the minor determinant of $X$ taking
$i$, $j$ and $k$th columns.
We have an action of the symmetric group $S_{10}$ which act as
a permutation of the columns of $X$. In terms of the coordinates
$(u_i,v_i)$, $i=5,\cdots,10$ the $S_{10}$-action can be written
as follows \cite{OD}:

The actions of $s_1, s_2, s_3$ are given by
\be\label{eq:sxy1}
\begin{array}{ll}
s_1(u_i)=\dfrac{1}{u_i}, &
s_1(v_i)=\dfrac{v_i}{u_i},\\[4mm]
s_2(u_i)=v_i, &
s_2(v_i)=u_i,\\[4mm]
s_3(u_i)=\dfrac{u_i-v_i}{1-v_i}, &
s_3(v_i)=\dfrac{v_i}{v_i-1},
\end{array}
\ee
The action of $s_4$ is
\be\label{eq:sxy2}
\begin{array}{llll}
s_4(u_5)=\dfrac{1}{u_5}, &
s_4(v_5)=\dfrac{1}{v_5}, &
s_4(u_i)=\dfrac{u_i}{u_5}, &
s_4(v_i)=\dfrac{v_i}{v_5},\quad (i=6,\ldots,10)
\end{array}
\ee
And $s_i$ for $i=5,\ldots,9$ act as
\be\label{eq:sxy3}
\begin{array}{llll}
s_i(u_i)=u_{i+1}, &
s_i(u_{i+1})=u_i, &
s_i(u_j)=u_j, & (j \neq i, i+1) \\
s_i(v_i)=v_{i+1}, &
s_i(v_{i+1})=v_i, &
s_i(v_j)=v_j, & (j \neq i, i+1)
\end{array}
\ee

Besides the permutations $s_i \in S_{10}$ ($i=1,\ldots,9$),
there exist another
important involution $s_0$ on the variables $(u_i, v_i)$, namely
\be\label{eq:sxy4}
s_0(u_i)=\dfrac{1}{u_i}, \quad
s_0(v_i)=\dfrac{1}{v_i}.
\ee
Geometrically, this is a standard Cremona transformation with
center $(P_1,P_2,P_3)$.
By direct computation, we have
\be
(s_0s_i)^2=1, \quad (i \neq 3) \quad
{\rm and} \quad (s_0s_3)^3=1.
\ee
In summary, the transformations $s_i$, $i=0,1,\ldots,9$ defined by
(\ref{eq:sxy1}), (\ref{eq:sxy2}), (\ref{eq:sxy3}) and
(\ref{eq:sxy4})
give a birational representation of the
Weyl group $W(E_{10})$ on the field of rational functions
${\bf{C}}(u_5,\ldots,u_{10},v_5,\ldots,v_{10})$.

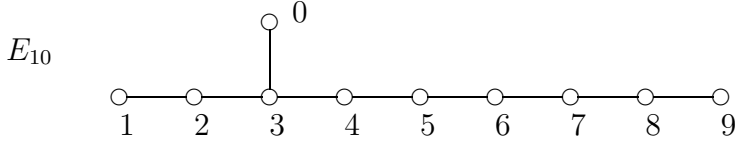
\setlength{\unitlength}{1mm}
\begin{picture}(100,25)(-20,-8)
\put(0,-5){1}
\put(10,-5){2}
\put(20,-5){3}
\put(30,-5){4}
\put(40,-5){5}
\put(50,-5){6}
\put(60,-5){7}
\put(70,-5){8}
\put(80,-5){9}
\put(23,10){0}
\put(-15,5){$E_{10}$}
\put(0,0){\circle{2}}
\put(10,0){\circle{2}}
\put(20,0){\circle{2}}
\put(30,0){\circle{2}}
\put(40,0){\circle{2}}
\put(50,0){\circle{2}}
\put(60,0){\circle{2}}
\put(70,0){\circle{2}}
\put(80,0){\circle{2}}
\put(20,10){\circle{2}}
\put(1,0){\line(1,0){8}}
\put(11,0){\line(1,0){8}}
\put(21,0){\line(1,0){8}}
\put(31,0){\line(1,0){8}}
\put(41,0){\line(1,0){8}}
\put(51,0){\line(1,0){8}}
\put(61,0){\line(1,0){8}}
\put(71,0){\line(1,0){8}}
\put(20,1){\line(0,1){8}}
\end{picture}

The construction of the elliptic Painlev\'e equation
is very simple.
The Weyl group $W(E_{10})$ contains
$W(E^{(1)}_{8})$ generated by $s_i$ ($i=0,\ldots,8$).
This group $W(E^{(1)}_{8})$ has a translation subgroup ${\bf Z}^8$.
The birational action of these translations on ${\cal M}$ is
nothing but the Sakai's elliptic Painlev\'e equation.
The explicit action of these translations on
the variables $(u_i,v_i)$ are too complicated
and seems to be beyond our computational ability.
We give an intermediate formula for one of the translations\footnote{
There exist $240$ commuting translations corresponding to $E_8$ roots.
Any of them can be represented as a composition of 58 simple reflections.
Among the $240$ translations, only 8 of them are
multiplicatively independent.}
\be\label{trT}
T=(pqp)^2, \quad
p=s_3s_4s_5 s_2s_3s_4 s_1s_2s_3 s_0,\quad
q=s_6s_7s_8 s_5s_6s_7 s_4s_5s_6.
\ee
The result is given as follows:
\be
\begin{array}l
p(u_5,u_6,u_i)=
\dfrac{\mu_{146}}{\mu_{156}}
\left(\dfrac{\mu_{256}}{\mu_{246}},
\dfrac{\mu_{356}}{\mu_{346}},
\dfrac{\mu_{56i}}{\mu_{46i}}\right), \quad (i=7,\ldots,10)\\[4mm]
p(v_5,v_6,v_i)=
\dfrac{\mu_{145}}{\mu_{156}}
\left(\dfrac{\mu_{256}}{\mu_{245}},
\dfrac{\mu_{356}}{\mu_{345}},
\dfrac{\mu_{56i}}{\mu_{45i}}\right),\quad (i=7,\ldots,10)
\end{array}
\end{equation}
\be
\begin{array}l
q(u_5,u_6,u_7,u_8,u_9,u_{10})=\dfrac{1}{u_7}
(u_8,u_9,1,u_5,u_6,u_{10}),\\[2mm]
q(v_5,v_6,v_7,v_8,v_9,v_{10})=\dfrac{1}{v_7}
(v_8,v_9,1,v_5,v_6,v_{10}).\\
\end{array}
\end{equation}

If the $9$ points $P_1, \ldots, P_9$ are in general position,
there exist unique elliptic curve $C \subset {\bf P}^2$
which pass through the $9$ points.
This curve $C$ play the role of the fiber at $u=\infty$ in the
previous section and it is invariant
under the action of $W(E^{(1)}_8)$.
Using this curve $C$ as a ``ruler'', Sakai introduced another coordinates
of the coset ${\cal M}$ :
$\theta_1,\ldots,\theta_9$, $\tau$ and $(x:y:z) \in {\bf P}^2$,
such that the matrix $X$ is represented as
\be
X=\left[\begin{array}{cccccccc}
\wp(\theta_1)&\wp(\theta_2)&\cdots&\wp(\theta_9)&x\\
\wp'(\theta_1)&\wp'(\theta_2)&\cdots&\wp'(\theta_9)&y\\
1&1&\cdots&1&z\\
\end{array}\right].
\ee
Here $\wp(\theta)=\wp(\theta,\tau)$ is the Weierstrass $\wp$ function
which parameterize the elliptic curve $C$.
In terms of Sakai's coordinates, the action of $W(E^{(1)}_8)$ is given as
follows \cite{Sakai}.
The $S_9$ part is just the permutation of the parameters $\theta_i$.
The only non-trivial one is $s_0$ which has been determined
explicitly\footnote{This means that these nine $\theta_i$'s transform
under $W(E_8)$ as the $SL(9)$ Cartan subalgebra, and hence  correspond to
the nine
radii in the $T^9$ compactification of $M$-theory.  The extra Weyl
reflection $s_0$
(called `2/5 transformation' in \cite{BFM}) is naturally understood via the
$SL(9)$ decomposition of $E_8$ \cite{Miz}.} as
\be
s_0(\theta_i)=\theta'_i=\left\{
\begin{array}{ll}
\theta_i+\dfrac{1}{3} (\theta_1+\theta_2+\theta_3),& i = 4,\ldots,9\\[4mm]
\theta_i-\dfrac{2}{3} (\theta_1+\theta_2+\theta_3),& i = 1,2,3
\end{array}
\right.
\ee
\be
s_0
\left[\begin{array}{l}
x\\y\\z
\end{array}\right]=
\left[\begin{array}{ccc}
x'_1&x'_2&x'_3\\
y'_1&y'_2&y'_3\\
z'_1&z'_2&z'_3
\end{array}\right]
\left[\begin{array}{l}
d_{23} l_{31}l_{12}\\d_{31} l_{12}l_{23}\\d_{12} l_{23}l_{31}
\end{array}\right],
\ee
\be
l_{jk}=\det
\left[\begin{array}{ccc}
x&x_j&x_k\\
y&y_j&y_k\\
z&z_j&z_k
\end{array}\right],\quad
d_{jk}=
\det\left[\begin{array}{ccc}
x_{\ast}&x_j&x_k\\
y_{\ast}&y_j&y_k\\
z_{\ast}&z_j&z_k
\end{array}\right]
\det\left[\begin{array}{ccc}
x'_{\ast}&x'_j&x'_k\\
y'_{\ast}&y'_j&y'_k\\
z'_{\ast}&z'_j&z'_k
\end{array}\right].
\ee
Where $(x_{\ast},y_{\ast},z_{\ast})$ and
$(x'_{\ast},y'_{\ast},z'_{\ast})$ are any points on the curve
$C$ such that
$(x_{\ast},y_{\ast},z_{\ast})=(\wp(\theta),\wp'(\theta),1)$,
$(x'_{\ast},y'_{\ast},z'_{\ast})=(\wp(\theta'),\wp'(\theta'),1)$ with
$\theta'=\theta+(\theta_1+\theta_2+\theta_3)/3$.

In these coordinates, the translation $T$ (\ref{trT}) acts on the parameters
$\theta_i$ as
\be\label{2/5like}
T(\theta_1,\ldots,\theta_9)=
(\theta_1,\ldots,\theta_9)-\dfrac{1}{3} \theta
(2,2,2,-1,\cdots,-1),
\ee
where $\displaystyle \theta=\sum_{i=1}^9 \theta_i$.
When $\theta=0$ (modulo periods), the first 9 points are in special position
such that the curve $C$ passing through the 9 points is given by one
parameter
family (a `pencil' of cubic)
\be
\lambda F(x,y,z)+\mu G(x,y,z)=0.
\ee
Then the corresponding del Pezzo $B_9$ admits an elliptic fibration
$B_9 \rightarrow {\bf P}^1=\{(\lambda:\mu)\}$ and
9 blown-up ${\bf P}^1$'s correspond to 9 sections of the fibration.
The parameters $\theta_i$ specify the 9 points
$(x_i:y_i:z_i)=(\wp(\theta_i):\wp'(\theta_i):1)$
where the sections intersect with the marked curve $C$ (at $u=\infty$).
These data define the Seiberg-Witten curve for $d=6$ $E_8$-string
as explained in \cite{Mo}.
In this special case, by choosing the parameter $(\lambda:\mu)$
suitably $C$ may pass the 10th point $(x:y:z)$ also and we can put
$(x:y:z)=(\wp(\theta_{10}):\wp'(\theta_{10}):1)$. Then all the action of
$W(E_{10})$ are represented by addition and permutation on the variables
$\theta_i$ ($i=1,\ldots,10$).

\section{Painlev\'e B\"acklund transformations and $U$-duality}

\subsection{Relation to $M$-theory duality}
The del Pezzo surfaces $B_k$ play crucial role
in various context of string compactifications.
In a recent paper \cite{INV}, Iqbal, Neitzke and Vafa observed a
duality between $M$-theory on $T^k$ and del Pezzo surfaces $B_k$.
In this correspondence, the Weyl group part of the $U$-duality group
is identified with the Cremona isometry $W(E_k)$ for del Pezzo $B_k$.
As we have seen in the previous section, the Cremona isometry is
the origin of the B\"acklund transformation/discrete
time evolution of the elliptic Painlev\'e equation, it is natural to  expect
some relation between the Painlev\'e B\"acklund transformations and
$U$-duality. In fact, there exist an analogy between these two Weyl
group realizations. Namely the
permutation part of the duality  can be realized as a change of
the order of the compactifications \cite{OP}, correspondingly
the Weyl group symmetry of the $E_{10}$ Painlev\'e equation appears
as a change of the blowing down structure \cite{Sakai}.

The Painlev\'e difference equations reduce to the six Painlev\'e
differential equations in the continuum limit. The latter also possess
affine Weyl
group symmetries generated by the B\"acklund transformations. Thus it will
be
interesting to explore whether these differential equations have direct
connections
with the string equations of motion.

In general relativity, it has been known for some time that some static,
axisymmetric solutions of Einstein(-Maxwell)'s equation(s) obey
Painlev\'e differential equations \cite{Ernst->Painleve}-\cite{CW}.
For example, the
Ernst equation, the equation of motion for the scalars in the dimensionally
reduced $D=4$ pure gravity, reduce to the third or the fifth Painlev\'e
equation under certain assumptions.  Since $D=10$, type IIB scalar sigma
model is identical to that of the Ernst system, one can exploit the general
relativity result to find special IIB scalar solutions that obey Painlev\'e
equations.  

Let us consider a consistent truncation of type IIB supergravity
\begin{equation}
{\cal L} =\sqrt{-G^{(10)}}\left(
R^{(10)}-\dfrac{\partial_M\tau\partial^M\overline{\tau}}{2(\mbox{Im}\tau)^2}
\right),\label{IIBaction}
\end{equation}
where $\tau=C+ie^{-\Phi}$ with $C$ and $\Phi$ being the RR scalar and the
dilaton, respectively. We further adopt an ansatz that the
ten-dimensional Einstein-frame metric $G^{(10)}_{MN}$ is of the form
\begin{equation}
ds_{\rm IIB}^2=\lambda^2(dx^2+d\rho^2)+\rho^2d\phi^2+
(-dt^2+\sum_{i=1}^6dx_i^2),
\label{IIBmetric}
\end{equation}
and that $\lambda$ is a real function of $-\infty<x<\infty$, the coordinate
parallel to the symmetric axis, and the radial coordinate $\rho\geq 0$.
$\phi$ is the angle coordinate. We also assume that the complex potential
$\tau$ depends\footnote{
One may also trade $\rho$ for the time $t$ to discuss colliding string wave
solutions. See e.g.
\cite{DMM} for recent discussions and further references.} only on
$x$ and $\rho$. In this way, we get a two-dimensional system without
enlarging the
duality symmetry than
$SL(2,\mbox{\boldmath R})/U(1)$.

In fact, this truncation is equivalent to the dimensional reduction of $D=4$
pure gravity to $D=2$ with a four-dimensional metric
\begin{equation}
ds_{\rm
4D}^2=e^{\Phi}(\lambda^2(dx^2+d\rho^2)+\rho^2
d\phi^2)-e^{-\Phi}(dt+A_\phi d\phi)^2
\label{4Dmetric}
\end{equation}
with
\begin{equation}
\partial_\xi A_\phi = -i\rho e^{2\Phi}\partial_{\overline{\xi}}C,~~~~~
\xi\equiv x+i\rho.
\end{equation}
The equation of motion for $\tau$ is given by the Ernst
equation
\begin{equation}
e^{-\Phi}\delta^{\mu\nu}\partial_\mu(\rho\partial_\nu\tau)
=-i\rho\delta^{\mu\nu}\partial_\mu\tau\partial_\nu\tau,
\label{Ernst_eq}\end{equation}
where $x^\mu=(x,\rho)$. If $\tau$ is known, the conformal factor $\lambda$
is
consistently determined by integrating the first-order `Virasoro
constraint'. 
(See \cite{Nic2} for related technology.)

The metric ansatz (\ref{IIBmetric}) is close to that for the D7-brane
solutions
\cite{D7}, but the crucial difference is the appearance of $\rho^2$ in
$G_{\phi\phi}^{(10)}$. Owing to this explicit coordinate dependence (`the
Weyl
canonical coordinate'), $\tau$ cannot be holomorphic, and the solution does
not preserve supersymmetry.

\subsection{Painlev\'e III and $S$-duality}\label{P3andSduality}
To reduce (\ref{Ernst_eq}) to a Painlev\'e equation, we first switch from
the $SL(2,R)$ variable $\tau$ to the $SU(1,1)$ variable $F$, defined by
\cite{LM}
\begin{equation}
\tau=i\dfrac{1+F}{1-F}.
\end{equation}
In terms of $F$, the Ernst equation becomes
\begin{equation}
(1-F\overline{F})\delta^{\mu\nu}\partial_\mu(\rho\partial_\nu F)
=-2\rho\delta^{\mu\nu}\overline{F}\partial_\mu F\partial_\nu F.
\label{F_equation}
\end{equation}
We further assume the coordinate dependence of $F(x,\rho)$ as
\begin{equation}
F(x,\rho)=f(\rho)e^{i\omega x},
\label{F_ansatz}
\end{equation}
where $f(\rho)$ is a real function, and $\omega$ is a real constant. The
equation (\ref{F_equation}) reduces to
\begin{equation}
(f^2-1)(f''+\dfrac{f'}{\rho}-\omega^2f)
=2f(f'^2-\omega^2 f^2).\label{realf''equation}
\end{equation}
Here the prime denotes differentiation with respect to $\rho$.
Then by the replacement
\begin{equation}
y\equiv\dfrac{1+f}{1-f}
\end{equation}
we obtain 
\begin{equation}
y''=\dfrac{y'^2}y-\dfrac{y'}\rho+\dfrac{\omega^2}4(y^3-\dfrac 1y).
\end{equation}
This is Painlev\'e III
\begin{equation}
y''=\dfrac{y'^2}y-\dfrac{y'}\rho
+\dfrac{\alpha y^2+\beta}\rho
+\gamma y^3+\dfrac \delta y
\end{equation}
with special parameters
\begin{equation}
\alpha=\beta=0, \quad \gamma=-\delta=\dfrac{\omega^2}4.\label{P3parameters}
\end{equation}

Painlev\'e III (with generic parameters) is known to have a symmetry of
B\"acklund transformations isomorphic to the Weyl group of type
$(A_1 \oplus A_1)^{(1)}$ generated by three independent Weyl reflections
\cite{Sakai}. One of them is
\begin{equation}
y\mapsto \dfrac 1y,
\quad
(\alpha, \beta, \gamma, \delta) \mapsto
(-\beta, -\alpha,-\delta,-\gamma)
\end{equation}
which leaves the condition (\ref{P3parameters}) unchanged. Since this
implies 
$\tau\mapsto-1/\tau$,
we see that this B\"acklund transformation of Painlev\'e III precisely
corresponds to $S$-duality of IIB theory.
On the other hand, the second B\"acklund transformation is simply
\begin{equation}
y\mapsto -y, \quad \rho\mapsto -\rho,
\quad
(\alpha, \beta, \gamma, \delta) \mapsto
(\alpha, \beta, \gamma, \delta)
\end{equation}
It just flips the sign of $\tau$, and hence is a physically irrelevant
transformation. 
Finally, Painlev\'e III has yet another independent B\"acklund
transformation. It shifts the parameters $\alpha$, $\beta$ to nonzero
values,
and therefore the differential equation does not keep its form of what has
been reduced from the Ernst equation. Thus it does not correspond to
duality, either.   

We conclude this subsection with a remark on how the Geroch group
\cite{Geroch}
is related to the Painlev\'e B\"acklund transformations. An affine Lie group
symmetry of a two-dimensional reduced nonlinear sigma model is a general
phenomenon \cite{Julia1981}, and in the present system
(\ref{IIBaction})(\ref{IIBmetric}) the symmetry is $A_1^{(1)}$,
the Geroch group. So the natural question is: how does its Weyl group piece
act on
the Painlev\'e equation? The answer is as follows:
Among two independent Weyl reflections of the Geroch group, one is
manifestly
realized in the sigma model (\ref{IIBaction}) as $S$-duality; this is also a
symmetry of the Painlev\'e equation, as we have seen above. The other is
obtained
by conjugating with the Kramer-Neugebauer (KN) involution
\footnote{The image of (\ref{4Dmetric}) under the KN involution is nothing
but the four-dimensional piece of the dual $M$-theory metric
\cite{Schwarz}.}
\cite{KN,BM}; this Weyl reflection is not the symmetry of the Painlev\'e
equation
because the KN involution does not preserve the metric ansatz
(\ref{F_ansatz}).  
 
\subsection{Comments on Painlev\'e V}
The Ernst equation is also known to reduce to the {\it fifth}
Painlev\'e equation by using a different ansatz
\cite{LM}.
We again start from the equations
(\ref{F_equation})(\ref{F_ansatz}), but this time we allow $f(\rho)$ to take
complex values. $\omega$ is a real constant, as before. In this case, the
equation (\ref{realf''equation}) is replaced by
\begin{equation}
(f\overline{f}-1)(f''+\dfrac{f'}{\rho}-\omega^2f)
=2\overline{f}(f'^2-\omega^2 f^2).
\label{complexf''equation}
\end{equation}
Multiplying $\overline{f}$ and subtracting the complex conjugate, we find
an integral
\begin{equation}
\dfrac{\rho(\overline{f}f'-f\overline{f}')}
{(1-f\overline{f})^2}=ia,
\end{equation}
where $a$ is a real integration constant. Writing
\begin{equation}
f(\rho)=r(\rho)e^{iu(\rho)}
\end{equation}
in terms of two real functions $r(\rho)$ and $u(\rho)$, we may express $u'$
as
\begin{equation}
u'=\dfrac{a(1-r^2)^2}{2\rho r^2}.
\end{equation}
Plugging them into (\ref{complexf''equation}), we obtain a second order
differential equation of a single variable $r(\rho)$. After a short
calculation we find
\begin{eqnarray}
Y''&=&\left(\dfrac 1{2Y}+\dfrac 1{Y-1}\right)Y'^2
-\dfrac{Y'}\rho -\dfrac{a^2(Y-1)^2}{2\rho^2}\left(Y-\dfrac{1}{Y}\right)
+\dfrac{2\omega^2 Y(Y+1)}{Y-1}
\end{eqnarray}
for $Y\equiv r^2$. This is the fifth Painlev\'e equation with parameters
\begin{equation}
\alpha=-\dfrac{a^2}{2}, \quad \beta=\dfrac{a^2}{2},
\quad \gamma=0, \quad \delta=2 \omega^2
\end{equation}
in the standard notation.

The $S$-duality transformation $\tau\mapsto -1/\tau$ acts on $r$ as
$r\mapsto -r$, 
which leaves $Y$ invariant. Therefore, it does not correspond to any of the
B\"acklund transformations of Painlev\'e V, but is trivially realized.


\section{Conclusions and Discussions}

In this paper, we studied the Weyl group symmetries from the point of view
of
Seiberg-Witten theory, the elliptic Painlev\'e equation and duality symmetry
of
$M$/string theory.
The results are summarized as follows:
\begin{itemize}
\item We have clarified the special role of the fiber at $u=\infty$ of the
Seiberg-Witten curves. The mass parameters, on which the Weyl group such as
$W(E^{(1)}_8)$ acts as the flavor symmetry,  are identified with the points
where the sections intersect the fiber.
\item We have given a simple formulation of the elliptic Painlev\'e equation
in which the hidden $W(E_{10})$ symmetry is manifestly realized.
The Seiberg-Witten geometry appears as a special case of this, where 
the solutions reduce to the elliptic functions.
\item We have studied some Painlev\'e differential equations arising from
dimensionally reduced equations of motion of strings.
In some special case, the B\"acklund transformation of the Painlev\'e
equation can be identified with a duality symmetry of $M$/string theory.
\end{itemize}

A property of the singularity confinement is proposed as a
discrete analog of the Painlev\'e property \cite{RGH}.
The singularity confinement demands that a singularity depending on the
initial data disappears after finite iteration of the mapping
and the memory of initial data is recovered.
Of course, the $E_{10}$ Painlev\'e equation has this property.
On the other hand, in \cite{BFM,KOSST}
it is argued that in $M$-theory, the apparent
cosmological singularities can be resolved by the duality transformations.
This phenomenon may be considered as the Painlev\'e property.
It should be also noted that the hyperbolicity of $W(E_{10})$
is crucial for the chaotic behavior of the cosmological singularity
in $M$-theory \cite{DHJN}.
In view of this and the symmetry structure, it is natural to guess that
the elliptic $E_{10}$ Painlev\'e equation, which is chaotic but integrable
in some sense, may play some role in certain
effective dynamics of $M$-theory on $T^{10}$.

\medskip
\noindent
{\it Acknowledgments}

We would like to dedicate this article to the memory of Sung-Kil Yang
from whom we learned a lot of things through encouraging and enjoyable
collaborations and who is now greatly missed.
We would also like to thank M.~Noumi, J.~Maharana, T.~Masuda and T.~Tani
for valuable discussions.

\end{document}